\documentclass[traditabstract]{aa} 
\usepackage{graphicx}
\usepackage{txfonts}
\usepackage{url}
\usepackage{natbib}
\bibpunct{(}{)}{;}{a}{}{,} 
\usepackage[utf8]{inputenc}
\usepackage{subfigure}
\usepackage{verbatim} 
\begin{document}

   \title{Three-minute wave enhancement in the solar photosphere}
   \author{M. Stangalini$^{1}$, F. Giannattasio$^{1}$, D. Del Moro$^{1}$, F. Berrilli$^{1}$}
   \institute{$^{1}$Dipartimento di Fisica, Universit\'a di Roma "Tor Vergata"\\
              \email{marco.stangalini@roma2.infn.it}}

 
  \abstract
   {It is a well-known result that the power of five-minute oscillations is progressively reduced by magnetic fields in the solar photosphere. Many authors have pointed out that this fact could be due to a complex interaction of many processes: opacity effects, MHD mode conversion and intrinsic reduced acoustic emissivity in strong magnetic fields. While five-minute oscillations are the dominant component in the photosphere, it has been shown that chromospheric heights are in turn dominated by three-minute oscillations. Two main theories have been proposed to explain their presence based upon resonance filtering in the atmospheric cavity and non linear interactions. In this work we show, through the analysis of IBIS observations of a solar pore in the photospheric Fe I $617.3$ nm line, that three-minute waves are already present at the height of formation of this line and that their amplitude depends on the magnetic field strength and is strictly confined in the umbral region.}

   \authorrunning{M. Stangalini}

\maketitle

%

\section{Introduction}
Oscillatory phenomena are ubiquitous in and around magnetically active regions on the Sun. Since their discovery \citep{1969SoPh....7..351B}, waves in sunspots have been detected in many solar magnetic features at all spatial scales and at all heights in the atmosphere as small perturbations in intensity and velocity \citep{Bogdan15022006, 2009ApJ...692.1211C}. Still a fully understood picture about their excitation mechanism and their interaction with complex magnetic features is missing \citep{2009AIPC.1170..547K}. Many open questions are waiting to be answered. Among the many we have for example the role of the residual convection inside the umbra of sunspots and the power reduction of the oscillations in strong magnetic fields. For a detailed treatment of these and many other aspects about waves in sunspots we refer to \citet{2009ASPC..416...31K} and \citet{Bogdan15022006}. Recently \citet{simoniello2010} pointed out that the amplitude reduction may be consistent with the MHD mode conversion theory. \\
The study of wave generation and propagation in the Sun's atmosphere also provides valuable information about the atmospheric structure itself and its dynamics \citep{1992sto..work..261L, Bogdan15022006}.\\
High-spectral and high-spatial resolution ground-based and space-borne observations have supported improvements in the theoretical interpretation of waves in the solar atmosphere over the last decade. In this context, a very promising observational tool for the investigation of the propagation of waves is represented by the multi-line spectroscopy \citep{2002A&A...381..253B, 2006ApJ...648L.151J, 2009ApJ...692.1211C, 2010ApJ...722..131F} which allows the estimation of the phase lag of the waves between different layers in the solar atmosphere. In particular, \citet{2009ApJ...692.1211C} have shown that different magnetic regions show distinct power spectrum features as we move from the photosphere to the chromosphere. They have shown that while small magnetic elements are dominated by five-minute oscillations in both the photosphere and the chromosphere, large magnetic structures like pores and sunspots, on the contrary,  are still dominated by five-minute oscillations in the photosphere, but in the chromosphere, their power spectrum peak shows a clear shift toward higher frequencies (three-minute oscillations or equivalently $5$ mHz).\\
Many competing theories have been proposed to explain the presence of three-minute waves in sunspots. \citet{1981SvAL....7...25Z,2008SoPh..251..501Z} proposed a resonant atmospheric cavity to explain the multiple peaks in the power spectra. \citet{1993ApJ...402..721C} proposed in turn a theory based on  eigenoscillations of sunspots, while \citet{1991A&A...250..235F} have demonstrated that the presence of a cutoff frequency in a stratified atmosphere may easily provide a basic physical mechanism for the onset of the three-minute regime in the chromosphere. \\
\citet{2010ApJ...719..357F} have shown, by using a $3D$ simulations of wave propagation in a sunspot atmosphere, that in such a magnetic structure the three-minute amplification in the chromosphere, in fact, comes from the effect of the acoustic cutoff frequency. Waves below the cutoff frequency ($\nu_{c}=5.2$ mHz) are not allowed to upward propagate and are evanescent. On the other hand, higher frequency waves are free to propagate toward the chromosphere. During the propagation they increase their amplitude as a response to the rapid drop of the density and, therefore, they are amplified, dominating the power spectrum. In small magnetic elements the cutoff frequency is much smaller due to the radiative losses \citep{2008ApJ...676L..85K}, allowing the propagation of five-minute waves toward the chromosphere. The cutoff frequency can be also lowered through the so called ramp effect allowing the propagation of low frequency waves in inclined magnetic field environment \citep{2006ApJ...648L.151J}. \citet{  refId0} found, using the same IBIS data investigated in this work, that the amount of energy transferred toward the upper layers of the Sun is strongly dependent on the complexity of magnetic field geometry, through a combination of ramp effect and MHD mode conversion.\\ 
In addition to this, many authors reported suspicious high-frequency power enhancements in rings surrounding active regions, both in the photosphere \citep{1992ApJ...394L..65B} and in the chromosphere \citep{1992ApJ...392..739B}. Later it was suggested by \citet{2009A&A...506L...5K} that these "acoustic halos" are probably generated by fast MHD mode refraction in the vicinity of the conversion layer (i.e. the layer where the sound speed equals the Alfven speed). More recently, \citet{2010SoPh..tmp...76S} pointed out that the halos of high frequency acoustic power are strongest at intermediate magnetic field strength ($150-250$ G), consistently with \citet{2002A&A...387.1092J} and  \citet{1998ApJ...504.1029H}, and they also discovered a clear association with horizontal magnetic fields. In addition, they also found that the frequency peak of the acoustic enhancement also slightly increases with the magnetic field strength. \\
In this work, we report on the presence of the three-minute waves also in the umbral photosphere of a pore. More in detail, we study the three-minute power enhancement and its relation to the magnetic field strength. By using a wavelet analysis, we also found that enhanced three-minute signal is non-stationary. The three-minute enhancement, moreover, interests only the umbra of the pore, as seen in intensity images, and inside this region the oscillations are in phase.
    \begin{figure*}[t]
   \centering
   \subfigure[Magnetogram]{\includegraphics		[width=5cm,trim=12mm 3mm 12mm 3mm, clip]{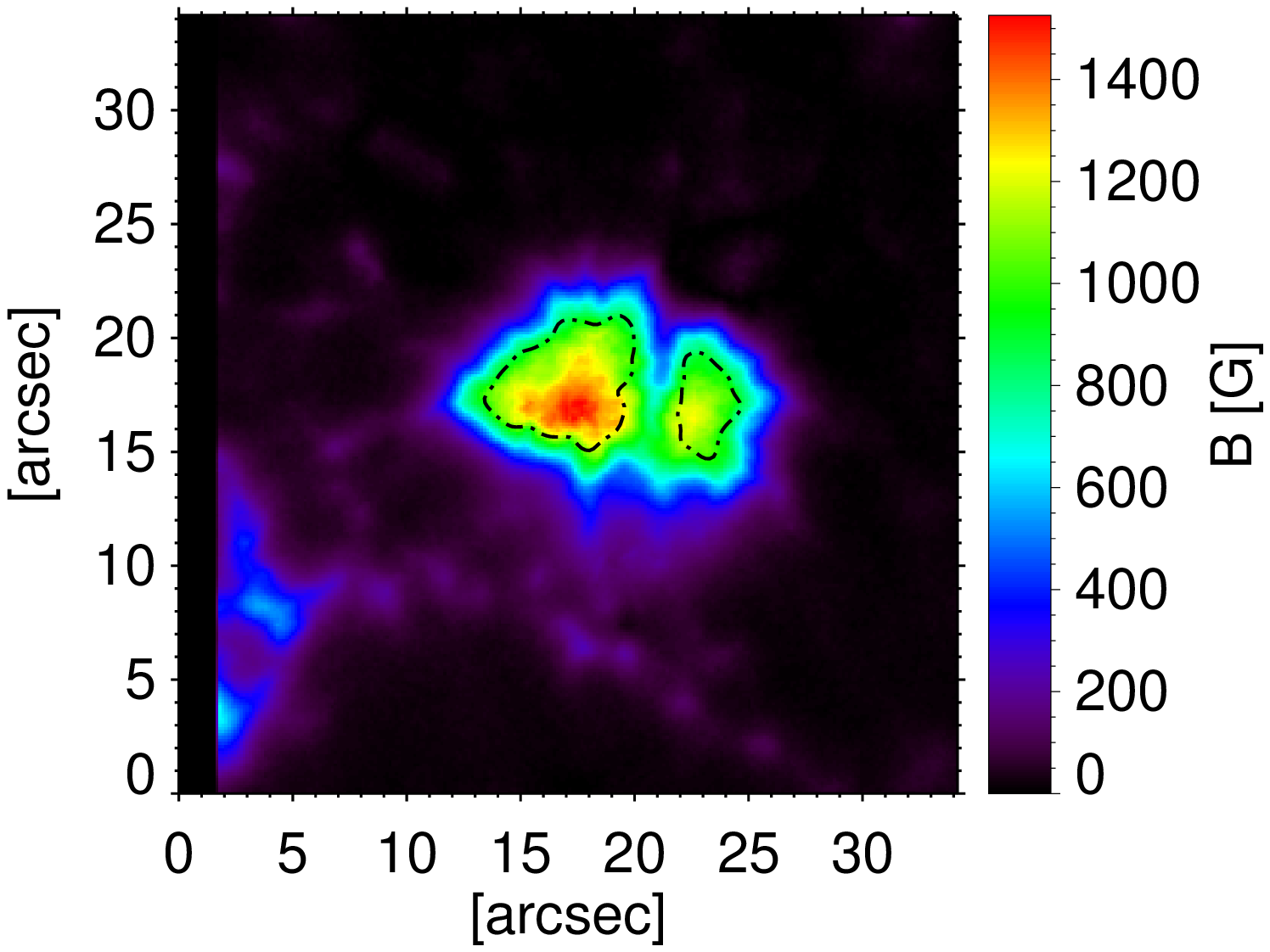}}
   \subfigure[Amplitude @ $3$ mHz]{\includegraphics	[width=5cm,trim=12mm 3mm 12mm 3mm, clip]{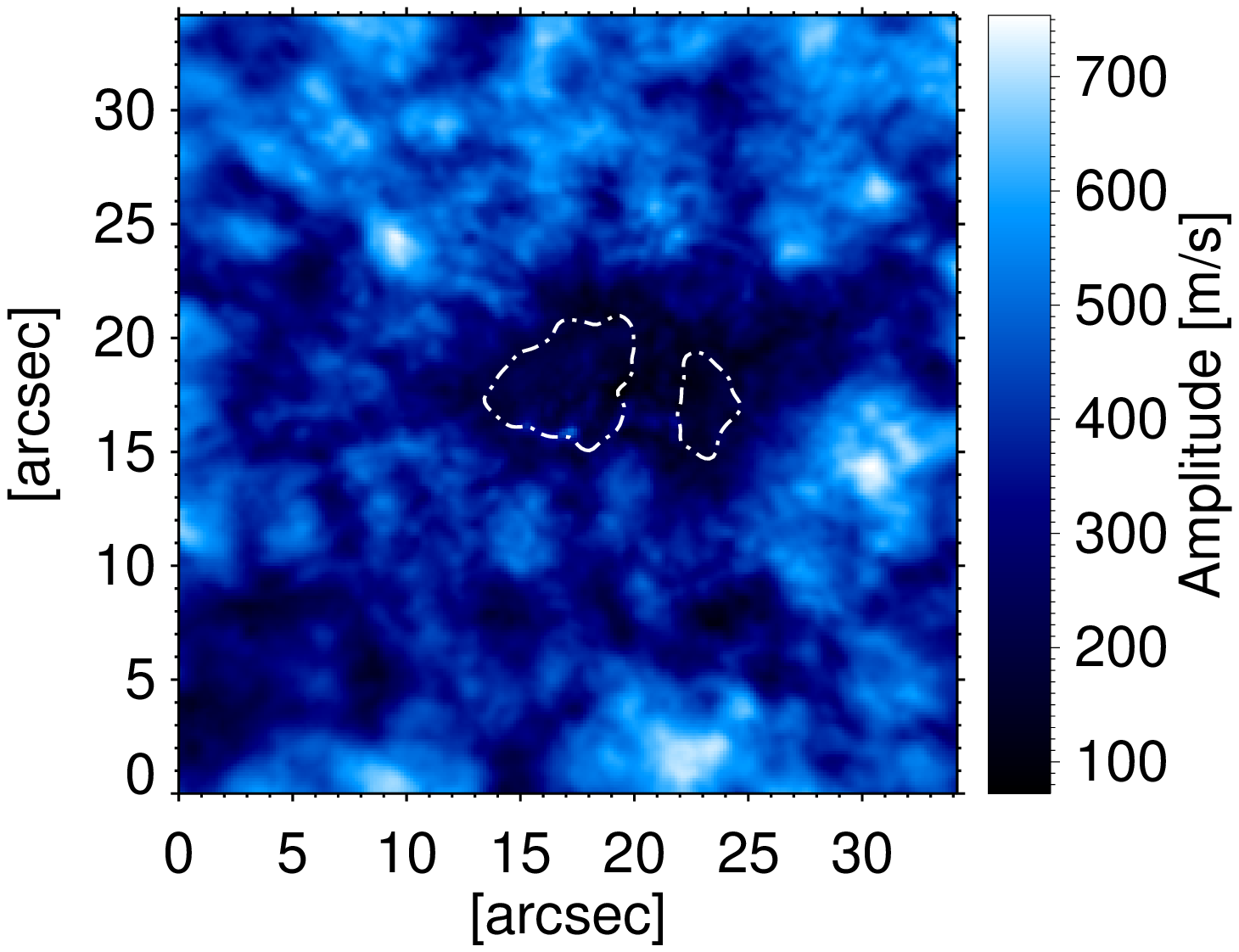}}
   \subfigure[Amplitute @ $5$ mHz]{\includegraphics	[width=5cm,trim=12mm 3mm 12mm 3mm, clip]{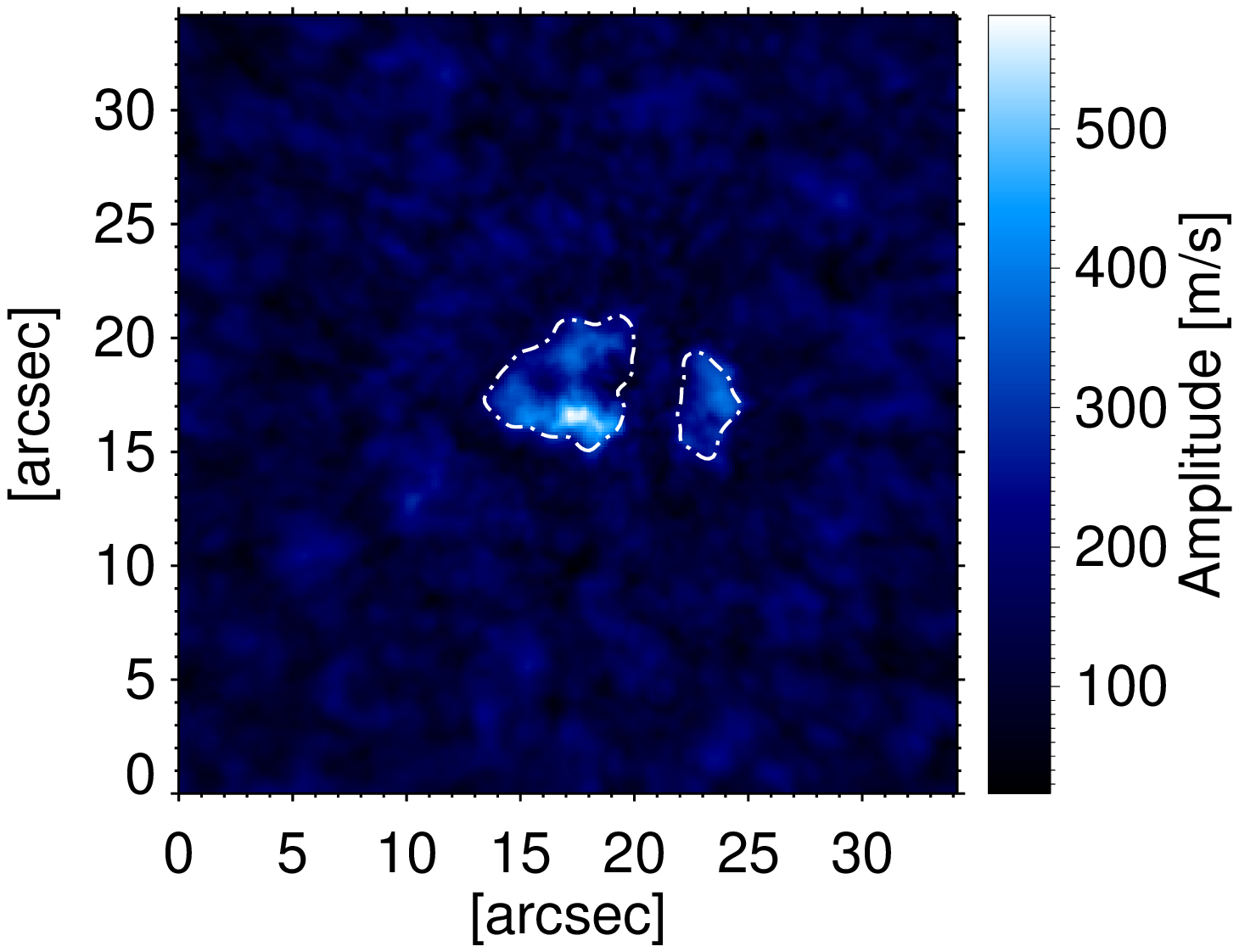}}
   \subfigure[Frequency peak]{\includegraphics		[width=5cm,trim=12mm 3mm 12mm 3mm, clip]{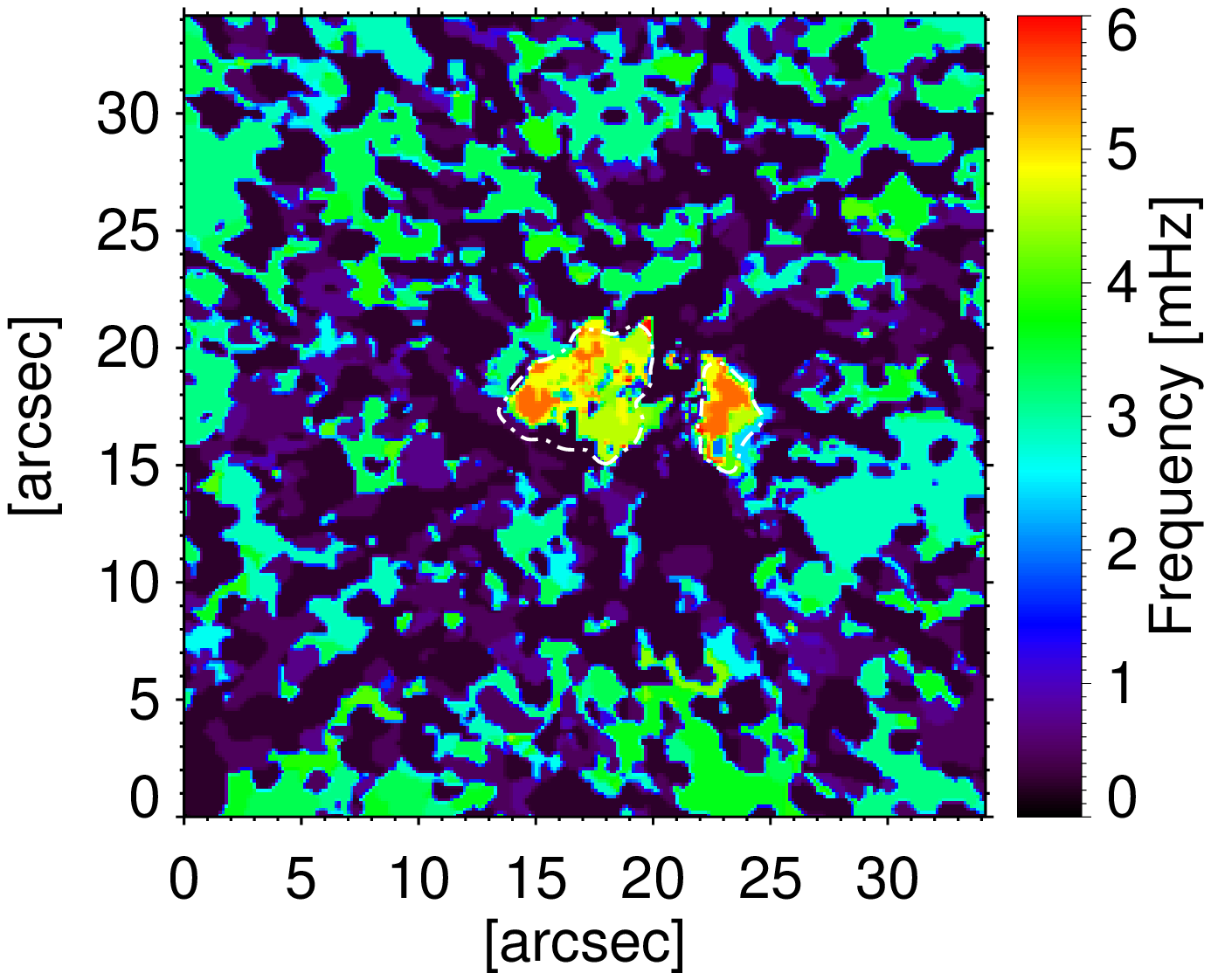}}
   \subfigure[Intensity core Fe $617.3$ nm]{\includegraphics[width=5cm,trim=12mm 3mm 12mm 3mm, clip]{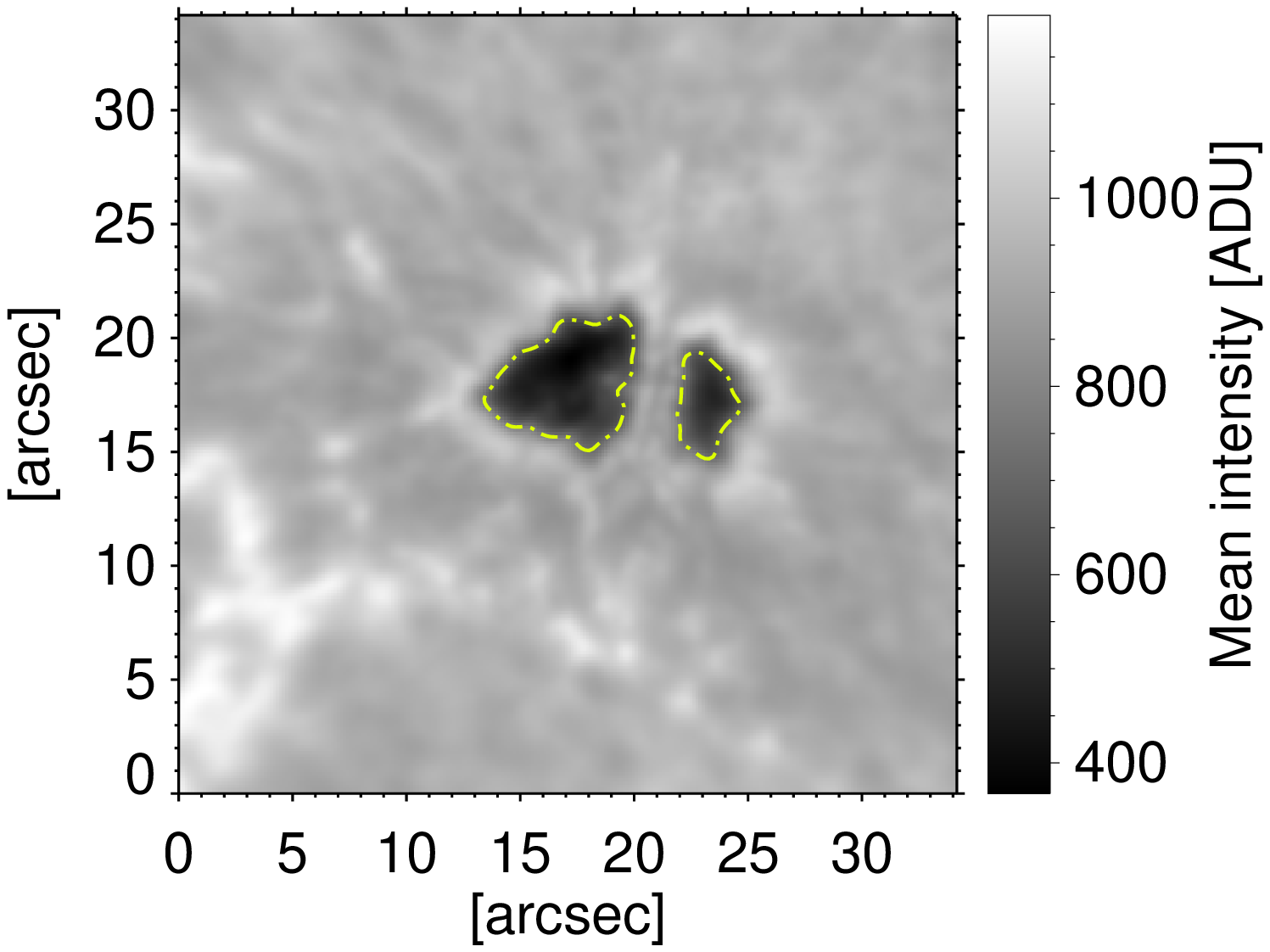}}

   \caption{(a) Center-of-gravity magnetogram Fe $617.3$ nm (absolute value). (b) Amplitude of $3$ mHz oscillations (Fe $617.3$ nm) (c) Amplitude of $5$ mHz oscillations (Fe $617.3$ nm). (d) Position in frequency of the power spectrum peak (Fe $617.3$ nm). The contour indicates the position of the umbra as seen in intensity images.} 
    \label{panels}
   \end{figure*}
\section{Observations}  
The data used in the work have been also used in \citet{refId0} and were acquired on October 15th 2008 in full Stokes mode with IBIS at DST. IBIS is the Interferometric BIdimensional Spectrometer based on a dual Fabry-Perot system combining high-spectral resolution and large FoV (field of view), as well as the ability to measure the polarization \citep{MScavallini06}.\\
The region observed was the AR11005 which consists of a small pore with a light bridge in the northern hemisphere [25.2$^{\circ}$ N, 10.0$^{\circ}$ W]. The data set consists of $80$ sequences, containing a full Stokes $21$ points scan of the Fe $617.3~nm$ line.
The temporal sampling is $52$ seconds and the pixel scale of these $512 \times 512$ pixel images was set at $0.167~arcsec$.\\
For further details on the data set and the calibration pipeline refer to \citet{refId0} and \citet{2009ApJ...700L.145V} respectively.
\section{Results}
\subsection{Three-minute enhancement}
\begin{figure}[t]
   \centering
   \subfigure{\includegraphics[width=6cm,trim=5mm 22mm 5mm 3mm, clip]{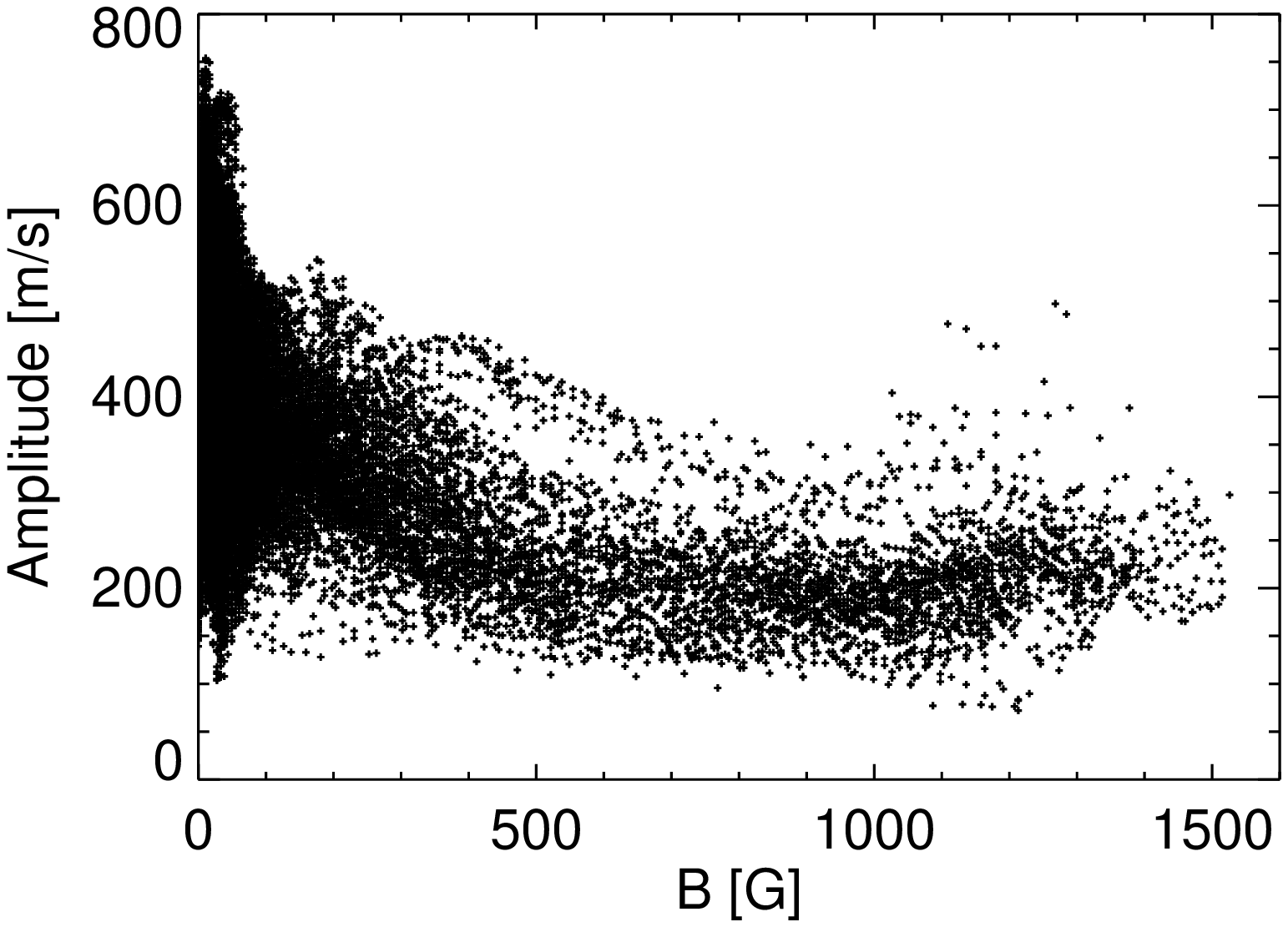}}
   \subfigure{\includegraphics[width=6cm,trim=5mm 3mm 5mm 10mm, clip]{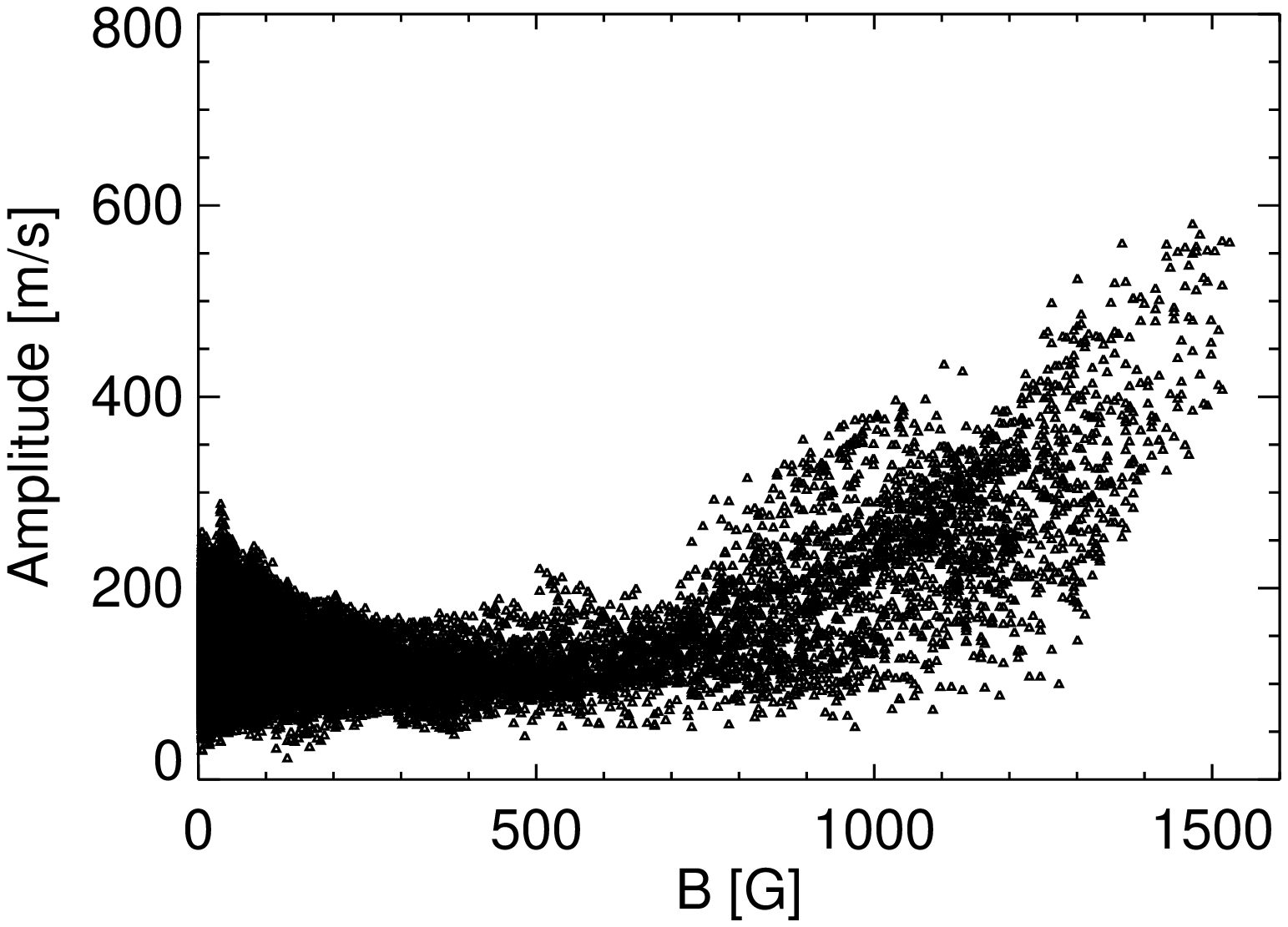}}
   \caption{(top) Amplitude of the five (upper panel) and three-minute (bottom panel) waves as a function of magnetic field strength. In this scatter plot each dot represents the amplitude of the wave and the magnetic field in one pixel in the spatial map shown in Fig. 1} 
    \label{trend}
   \end{figure}

In this work, we studied the amplitude enhancement of three-minute waves and the five-minute wave absorption as a function of the magnetic field strength. 
In order to limit the distortions of the FFT power spectrum estimation due to the limited length of the time series \citep{1970WRR.....6.1601E}, we used the Blackman-Tukey method with a Barlett windowing function \citep{blackman1958measurement} to estimate the amplitude spectrum.\\
We produced maps of the amplitude of the oscillations in two spectral bands, namely $3$ mHz and $5$ mHz, corresponding to periods of three and five minutes, by integrating over a $1$ mHz (panels c and b of Fig. \ref{panels}). In the following we will use the periods to refer to the respective oscillations.\\
We then studied the behaviour of the amplitude of the waves as a function of the magnetic field, estimated by means of the center of gravity (COG, see \citet{2003ApJ...592.1225U} for details) method (panel (a) of Fig. \ref{panels}).
Our power maps (panel (c) of Fig. \ref{panels}) reveal a power enhancement of the three-minute waves for magnetic field strength larger than $1000$ G. No acoustic halo is apparent in our maps.  
In Fig. \ref{trend} we show the behaviour of the amplitude of the oscillations in both spectral bands versus the magnetic field strength. The five-minute waves (upper panel) are absorbed as expected, and their amplitude is reduced roughly by a factor of $2$. Conversely, the three-minute oscillations (bottom panel) are largely enhanced when the magnetic field strength increases. The amplitudes of the three-minute waves is almost constant up to $700-1000$ G and then it suddenly increases for larger field strengths. \\
To further investigate this scenario we study the frequency shift of the power spectrum peak throughout the FoV. This analysis is reported in panel (d) of Fig. \ref{panels}.
In this map it is evident that in the quiet Sun, but also in the diffuse magnetic field surrounding the pore, the power spectrum peaks around $3$ mHz. Three-minute waves ($5$ mHz) are in turn the dominant component inside the umbra of the pore. For comparison panel (e) of Fig. \ref{panels} shows the intensity map in the core of the Fe $617.3$ nm spectral line.\\
It is worth noting that the frequency shift occurs abruptly at the boundaries of the umbra without any smooth transition from the low values observed outside ($3$ mHz) to the higher frequencies observed inside the pore ($5$ mHz).

\subsection{Wavelet analysis and spatial coherence}
\begin{figure}[]
\centering
   \subfigure{\includegraphics[width=6cm,trim=5mm 20mm 5mm 3mm, clip]{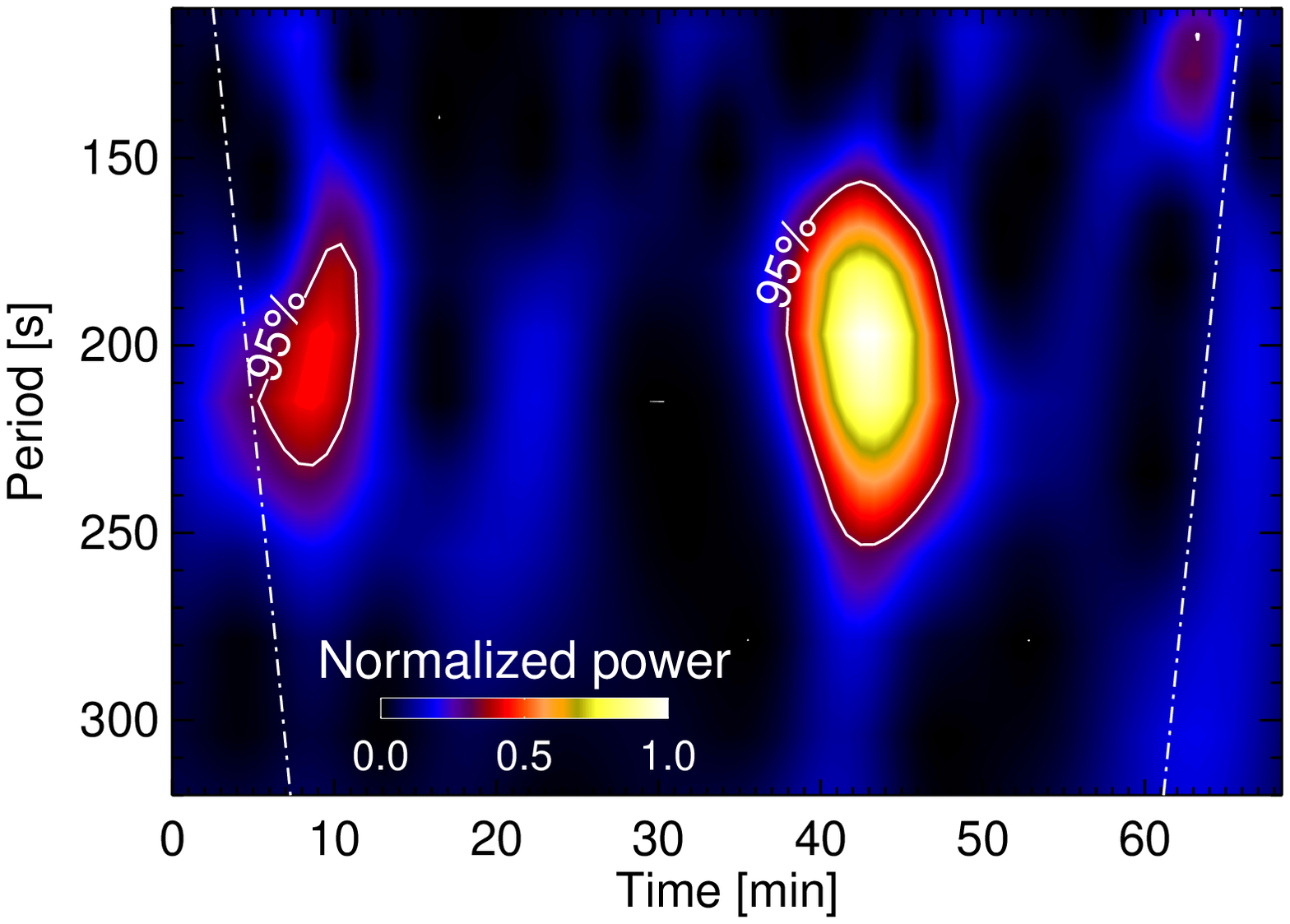}}
   \subfigure{\includegraphics[width=6cm,trim=5mm 3mm 5mm 10mm, clip]{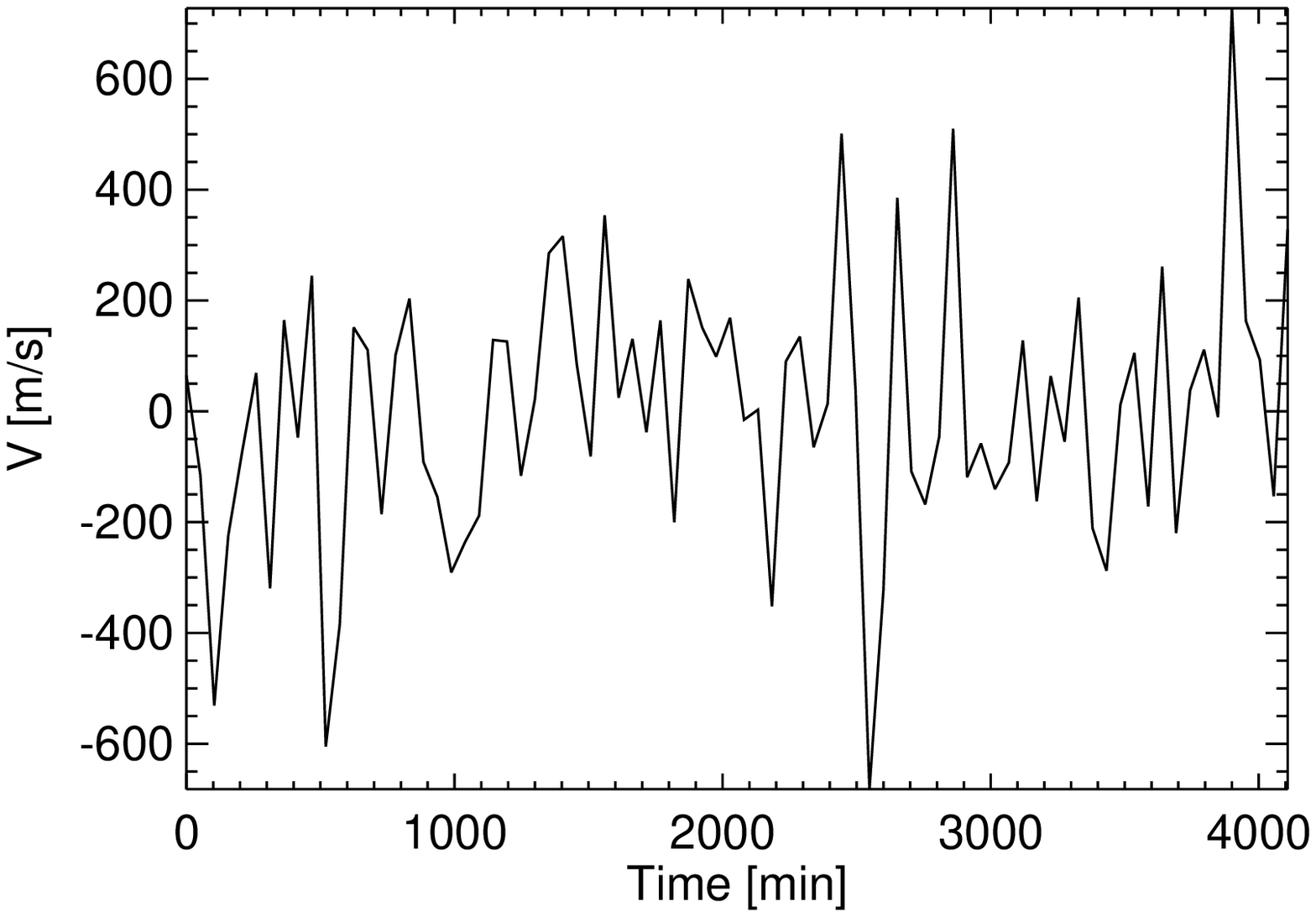}}
\caption{Wavelet spectrum (top panel) of the velocity signal averaged over the entire umbra (bottom panel). Positive values indicate redshifts. The continuous contours indicate the $95\%$ confidence level, dot-dashed contours indicate the cone of influence.}
\label{wavelet}
\end{figure}

\begin{figure}[t]
\centering
\includegraphics[width=9cm]{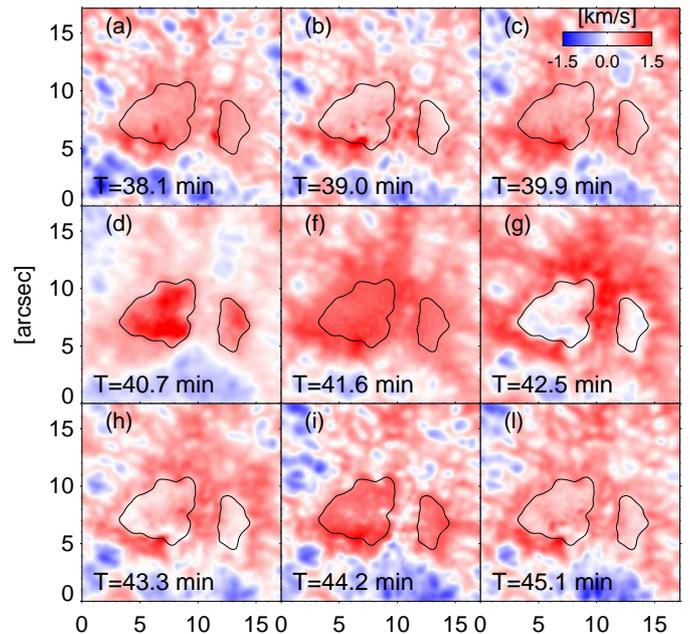}
\caption{Temporal evolution of the velocity field in the Fe $617.3$ nm. Positive values are redshifts.}
\label{temporal}
\end{figure}

To better analyse the nature of the three-minute signal in the umbra of the pore we used a wavelet analysis with a Morlet mother function. We studied the spatially averaged signal in the umbra delineated by the contours in Fig. \ref{panels}. In Fig. \ref{wavelet} it is shown the wavelet spectrum (upper panel) and the spatially averaged signal (bottom panel). The spectrum shows at least two main features located in time at $10$ min and $43$ min where the signal manifests three-minute oscillations (over $95\%$ confidence level). The waiting time between these two wave trains is around $30$ min, even though the shortness of our data does not allow us to provide a statistically relevant estimate. Interestingly, these non-stationary oscillations are spatially coherent throughout the umbra of the structure. This is apparent in Fig. \ref{temporal} where we show the temporal evolution of the velocity field during the onset of the second power peak shown in the wavelet spectrum ($38~min<t<45~min$). All the spatial positions within the umbra oscillate in phase, this is evident in panels (d) and (g) of Fig. \ref{temporal}.
Another interesting feature which is worth noting is visible in panels (d-e). After the rapid increase of the velocity (redshift) inside the umbra, the surrounding region around the umbra itself appears to be influenced by a velocity perturbation too, which resembles an expanding wave.

\section{Discussion}
The results shown provide clues on the three-minute waves generation in active regions by finding a clear three-minute enhancement in a photospheric magnetic region. Three-minute waves are generally expected to be dominant at chromospheric heights in large magnetic elements \citep{2009ApJ...692.1211C}.
In this work we have shown that the three-minute wave enhancement is confined in the region within the umbra of the pore, as seen in intensity images. This boundary is not coincident with the boundaries of the magnetic signal, which extends a few arcseconds beyond the umbra. As a consequence, the three-minute enhancement appears to be determined by a different regime of the magneto-convection. As we go into the umbra boundaries we observe an abrupt increase of the three-minute power as we can see in the frequency map (panel (d) of Fig. \ref{panels}). The frequency shift matches exactly the umbra boundary, demonstrating the onset of a different regime for the oscillation field.\\
Whatever the sources of the three-minute oscillations origin from, it appears that they are strongly dependent on the magneto-convection regime, and our results firmly point to this.  \\
P-mode absorption in magnetic structures has been investigated in details by many authors. Among the many mechanisms proposed so far to explain this power suppression, at least two of them are worthwhile to comment in light of our results: the opacity effect and the local suppression due to reduced emissivity. 
As for the first one, it is a well known result that the magnetic field, exerting its own pressure, empties the plasma environment (Wilson depression). This means that the spectral line used is forming well below the height of formation reached in the quiet Sun, allowing us to observe deeper and deeper in the solar atmosphere as the magnetic field increases. In addition to this, we have to keep in mind that due to the rapid drop of the density with height, the amplitude of the waves is subject to a rapid increase during their upward propagation. These two facts together mean that, in magnetic environments, we expect to see a suppression of the wave amplitude due to the Wilson depression. The latter mechanism also goes in the same direction producing a power suppression within magnetic regions. Both mechanisms should work indifferently for the five-minute waves as well as for three-minute waves. Our results suggest in turn that this cannot be the case and the three-minute waves, contrarily to what happens with the five-minutes, are enhanced in the umbra of the pore.\\ 
Moreover, \citet{1991A&A...250..235F} and  \citet{2010ApJ...719..357F} have argued that the three-minute waves may originate from a very basic physical effect due to the presence of the cutoff frequency. All the waves with frequency above the cutoff are free to upward propagate, increasing their amplitude, as we already said. On the other hand, waves with smaller frequencies are not free to propagate. This means that at higher layers the three-minute waves dominate the spectrum of oscillations.\\
Involving the cutoff frequency is appealing but the altitude of formation of the Fe $617.3$ nm spectral line used in this work ($300$ km in the quiet Sun \citep{2006ASPC..358..193N} and even below within the magnetic region) is not sufficient to explain the observed amplification. In addition we should see this effect also in other regions of the FoV (for example in the diffuse magnetic field).\\
By using a wavelet analysis, we also noted that the three-minute waves found in the umbra are not stationary and present wave-trains lasting for a few minutes (less than $10$ minutes). More interestingly the umbral oscillations show a clear spatial coherence involving the entire umbra. In addition, the two umbral regions split up by the presence of a light bridge also present oscillations in phase, probably indicating a generation mechanism of the three-minute waves involving the structure as a whole.

\section{Conclusions}
In this work we have reported on a three-minute wave enhancement in the magnetic umbra of a pore at photospheric heights. Making use of IBIS observations we have shown that the three-minute enhancement is strictly confined in the umbral region. This means that it is not dependent on the magnetic field alone but also on the magneto-convection regime. This seems to be in contrast with the behaviour of the five-minute waves which are in turn suppressed as expected. In addition, we have shown that the three-minute oscillations are strictly in phase throughout the umbra even if it is split into two part by the presence of a light bridge.\\
The inspection of the velocity fields, during the appearance of the three-minute wave-train, suggests the presence of an expanding wavefront produced by the perturbations within the umbra. This wavefront appears to be the echo of the three-minute oscillations which are enhanced inside the umbra.\\   
This observational results, as far as we know,  has never been reported so far, and we believe that further multi-height observations are required to provide a fairly consistent picture explaining this interesting three-minute enhancing mechanisms.

\begin{acknowledgements}
We acknowledge Serena Criscuoli for providing the COG magnetograms. Wavelet software was provided by C. Torrence and G. Compo, and is available at URL: http://atoc.colorado.edu/research/wavelets/.\\
IBIS was built by INAF-Osservatorio Astrofisico di Arcetri with contributions from the Università di Firenze and the Università di Roma "Tor Vergata". 
\end{acknowledgements}

\bibliography{references}

\begin{thebibliography}{29}
\expandafter\ifx\csname natexlab\endcsname\relax\def\natexlab#1{#1}\fi

\bibitem[{{Beckers} \& {Tallant}(1969)}]{1969SoPh....7..351B}
{Beckers}, J.~M. \& {Tallant}, P.~E. 1969, Sol. Phys., 7, 351

\bibitem[{{Berrilli} {et~al.}(2002){Berrilli}, {Consolini}, {Pietropaolo},
  {Caccin}, {Penza}, \& {Lepreti}}]{2002A&A...381..253B}
{Berrilli}, F., {Consolini}, G., {Pietropaolo}, E., {et~al.} 2002, A\&A, 381,
  253

\bibitem[{Blackman \& Tukey(1958)}]{blackman1958measurement}
Blackman, R. \& Tukey, J. 1958, The measurement of power spectra from the point
  of view of communications (Dover Publications)

\bibitem[{Bogdan \& Judge(2006)}]{Bogdan15022006}
Bogdan, T. \& Judge, P. 2006, Philosophical Transactions of the Royal Society
  A: Mathematical, Physical and Engineering Sciences, 364, 313

\bibitem[{{Braun} {et~al.}(1992){Braun}, {Lindsey}, {Fan}, \&
  {Jefferies}}]{1992ApJ...392..739B}
{Braun}, D.~C., {Lindsey}, C., {Fan}, Y., \& {Jefferies}, S.~M. 1992, \apj,
  392, 739

\bibitem[{{Brown} {et~al.}(1992){Brown}, {Bogdan}, {Lites}, \&
  {Thomas}}]{1992ApJ...394L..65B}
{Brown}, T.~M., {Bogdan}, T.~J., {Lites}, B.~W., \& {Thomas}, J.~H. 1992, ApJL,
  394, L65

\bibitem[{{Cally} \& {Bogdan}(1993)}]{1993ApJ...402..721C}
{Cally}, P.~S. \& {Bogdan}, T.~J. 1993, \apj, 402, 721

\bibitem[{{Cavallini}(2006)}]{MScavallini06}
{Cavallini}, F. 2006, Sol. Phys., 236, 415

\bibitem[{{Centeno} {et~al.}(2009){Centeno}, {Collados}, \& {Trujillo
  Bueno}}]{2009ApJ...692.1211C}
{Centeno}, R., {Collados}, M., \& {Trujillo Bueno}, J. 2009, ApJ, 692, 1211

\bibitem[{{Edge} \& {Liu}(1970)}]{1970WRR.....6.1601E}
{Edge}, B.~L. \& {Liu}, P.~C. 1970, Water Resources Research, 6, 1601

\bibitem[{{Felipe} {et~al.}(2010{\natexlab{a}}){Felipe}, {Khomenko}, \&
  {Collados}}]{2010ApJ...719..357F}
{Felipe}, T., {Khomenko}, E., \& {Collados}, M. 2010{\natexlab{a}}, ApJ, 719,
  357

\bibitem[{{Felipe} {et~al.}(2010{\natexlab{b}}){Felipe}, {Khomenko},
  {Collados}, \& {Beck}}]{2010ApJ...722..131F}
{Felipe}, T., {Khomenko}, E., {Collados}, M., \& {Beck}, C. 2010{\natexlab{b}},
  ApJ, 722, 131

\bibitem[{{Fleck} \& {Schmitz}(1991)}]{1991A&A...250..235F}
{Fleck}, B. \& {Schmitz}, F. 1991, \aap, 250, 235

\bibitem[{{Hindman} \& {Brown}(1998)}]{1998ApJ...504.1029H}
{Hindman}, B.~W. \& {Brown}, T.~M. 1998, ApJ, 504, 1029

\bibitem[{{Jain} \& {Haber}(2002)}]{2002A&A...387.1092J}
{Jain}, R. \& {Haber}, D. 2002, A\&A, 387, 1092

\bibitem[{{Jefferies} {et~al.}(2006){Jefferies}, {McIntosh}, {Armstrong},
  {Bogdan}, {Cacciani}, \& {Fleck}}]{2006ApJ...648L.151J}
{Jefferies}, S.~M., {McIntosh}, S.~W., {Armstrong}, J.~D., {et~al.} 2006, ApJL,
  648, L151

\bibitem[{{Khomenko}(2009)}]{2009ASPC..416...31K}
{Khomenko}, E. 2009, in Astronomical Society of the Pacific Conference Series,
  Vol. 416, Solar-Stellar Dynamos as Revealed by Helio- and Asteroseismology:
  GONG 2008/SOHO 21, ed. {M.~Dikpati, T.~Arentoft, I.~Gonz{\'a}lez
  Hern{\'a}ndez, C.~Lindsey, \& F.~Hill}, 31

\bibitem[{{Khomenko} {et~al.}(2008){Khomenko}, {Centeno}, {Collados}, \&
  {Trujillo Bueno}}]{2008ApJ...676L..85K}
{Khomenko}, E., {Centeno}, R., {Collados}, M., \& {Trujillo Bueno}, J. 2008,
  ApJL, 676, L85

\bibitem[{{Khomenko} \& {Collados}(2009)}]{2009A&A...506L...5K}
{Khomenko}, E. \& {Collados}, M. 2009, A\&A, 506, L5

\bibitem[{{Kosovichev}(2009)}]{2009AIPC.1170..547K}
{Kosovichev}, A.~G. 2009, in American Institute of Physics Conference Series,
  Vol. 1170, American Institute of Physics Conference Series, ed. {J.~A.~Guzik
  \& P.~A.~Bradley}, 547--559

\bibitem[{{Lites}(1992)}]{1992sto..work..261L}
{Lites}, B.~W. 1992, in NATO ASIC Proc. 375: Sunspots. Theory and Observations,
  ed. {J.~H.~Thomas \& N.~O.~Weiss}, 261--302

\bibitem[{{Norton} {et~al.}(2006){Norton}, {Pietarila Graham}, {Ulrich},
  {Schou}, {Tomczyk}, {Liu}, {Lites}, {L{\'o}pez Ariste}, {Bush},
  {Socas-Navarro}, \& {Scherrer}}]{2006ASPC..358..193N}
{Norton}, A.~A., {Pietarila Graham}, J.~D., {Ulrich}, R.~K., {et~al.} 2006, in
  Astronomical Society of the Pacific Conference Series, Vol. 358, Astronomical
  Society of the Pacific Conference Series, ed. {R.~Casini \& B.~W.~Lites},
  193--+

\bibitem[{{Schunker} \& {Braun}(2010)}]{2010SoPh..tmp...76S}
{Schunker}, H. \& {Braun}, D.~C. 2010, Sol. Phys., 76

\bibitem[{{Simoniello} {et~al.}(2010){Simoniello}, {Finsterle, W.},
  {Garc\'{\i}a, R. A.}, {Salabert, D.}, {Jim\'enez, A.}, {Elsworth, Y.}, \&
  {Schunker, H.}}]{simoniello2010}
{Simoniello}, R., {Finsterle, W.}, {Garc\'{\i}a, R. A.}, {et~al.} 2010, A\&A,
  516, A30

\bibitem[{{Stangalini} {et~al.}(2011){Stangalini}, {Del Moro, D.}, {Berrilli,
  F.}, \& {Jefferies, S. M.}}]{refId0}
{Stangalini}, M., {Del Moro, D.}, {Berrilli, F.}, \& {Jefferies, S. M.} 2011,
  A\&A, 534, A65

\bibitem[{{Uitenbroek}(2003)}]{2003ApJ...592.1225U}
{Uitenbroek}, H. 2003, ApJ, 592, 1225

\bibitem[{{Viticchi{\'e}} {et~al.}(2009){Viticchi{\'e}}, {Del Moro},
  {Berrilli}, {Bellot Rubio}, \& {Tritschler}}]{2009ApJ...700L.145V}
{Viticchi{\'e}}, B., {Del Moro}, D., {Berrilli}, F., {Bellot Rubio}, L., \&
  {Tritschler}, A. 2009, ApJL, 700, L145

\bibitem[{{Zhugzhda}(2008)}]{2008SoPh..251..501Z}
{Zhugzhda}, Y.~D. 2008, \solphys, 251, 501

\bibitem[{{Zhugzhda} \& {Locans}(1981)}]{1981SvAL....7...25Z}
{Zhugzhda}, Y.~D. \& {Locans}, V. 1981, Soviet Astronomy Letters, 7, 25

\end{thebibliography}
\bibliographystyle{aa}
\end{document}